\documentclass[]{article}
\usepackage{arxiv}
\usepackage{hyperref}
\usepackage{appendix}

\usepackage{siunitx}
\usepackage{moreverb}
\usepackage{graphicx}
\usepackage{latexsym}
\usepackage{url}
\usepackage{amssymb, amsmath, graphicx, bm, caption, verbatim, subcaption}
\usepackage{tikz,fancybox}
\usepackage[export]{adjustbox} 
\usepackage{array}
\usepackage{booktabs} 
\usepackage{cancel}

\definecolor{lightgray}{gray}{0.9}
\definecolor{groen}{rgb}{0.0, 0.5, 0.0}

\newcommand{\ctg}[1]{\textcolor{groen}{#1}}
\newcommand{\ctb}[1]{\textcolor{blue}{ {#1} }}

\newcommand{\cip}{\mbox{$\perp\!\!\!\perp$}}
\newcommand{\bbeta}{\mbox{\boldmath $\beta$}}
\newcommand{\bgamma}{\mbox{\boldmath $\gamma$}}
\newcommand{\bzeta}{\mbox{\boldmath $\zeta$}}

\newcolumntype{C}{>{\centering\arraybackslash}p{0.4cm}}

\newcommand{\splitcell}[1]{%
  \begin{tabular}{@{}l@{}}#1\end{tabular}%
}
\newcommand{\bigcheckmark}{\smash{\large\checkmark}}

\newcommand\BibTeX{{\rmfamily B\kern-.05em \textsc{i\kern-.025em b}\kern-.08em
T\kern-.1667em\lower.7ex\hbox{E}\kern-.125emX}}

\title{A framework for meta-analysis through standardized survival curves}

\author{ Joris Hautekiet \\ 
	 Unit of Cancer Epidemiology, Belgian Cancer Centre; Sciensano, Belgium \\
	 Department of Applied Mathematics, Computer Science and Statistics; Ghent University, Belgium \\
	 \AND
	 Marc Arbyn \\
	 Unit of Cancer Epidemiology, Belgian Cancer Centre; Sciensano, Belgium \\
	 \And
	Elena-Sophie Prigge \\
	Department of Applied Tumor Biology, Institute of Pathology, University of Heidelberg, \\
	Clinical Cooperation Unit Applied Tumor Biology; German Cancer Research Center (DKFZ), Germany \\
	\And
	Theresa Obermueller \\
	Department of Applied Tumor Biology, Institute of Pathology, University of Heidelberg, \\
	Clinical Cooperation Unit Applied Tumor Biology; German Cancer Research Center (DKFZ), Germany \\
	\And
	Els Goetghebeur \\
	Department of Applied Mathematics, Computer Science and Statistics; Ghent University, Belgium \\
	\texttt{els.goetghebeur@ugent.be} \\
}

\renewcommand{\headeright}{J. Hautekiet and E. Goetghebeur}
\renewcommand{\undertitle}{ }
\renewcommand{\shorttitle}{ }

\begin{document}

\maketitle


\begin{abstract}
Meta-analyses of survival studies aim to reveal the variation of an effect measure of interest over different studies and present a meaningful summary. They must address between study heterogeneity in several dimensions and eliminate spurious sources of variation. Forest plots of the usual (adjusted) hazard ratios are fraught with difficulties from this perspective since both the magnitude and interpretation of these hazard ratios depend on factors ancillary to the true study-specific exposure effect. 
These factors generally include the study duration, the censoring patterns within studies,  the covariates adjusted for and their distribution over exposure groups. Ignoring these mentioned features and accepting implausible hidden assumptions may critically affect interpretation of the pooled effect measure.

Risk differences or restricted mean effects over a common follow-up interval and balanced distribution of a covariate set are natural candidates for exposure evaluation and possible treatment choice. 
In this paper, we propose differently standardized survival curves over a fitting time horizon, targeting various estimands with their own transportability.  With each type of standardization comes a given interpretation within studies and overall, under stated assumptions. These curves can in turn be summarized by  standardized study-specific contrasts, including hazard ratios with more consistent meaning. We prefer forest plots of risk differences at well chosen time points. Our case study examines overall survival among anal squamous cell carcinoma patients, expressing the tumor marker $\text{p16}^{INK4a}$ or not, based on the individual patient data of six studies.
\end{abstract}


\section{Introduction}

Personalized medicine demands evidence on the effect of ever more detailed information or biomarkers on outcome. This must be obtained from sufficiently rich or large data bases, which may be found in randomized clinical trials (RCT), disease registers or local hospital based studies. In all instances an appreciation of between and within study heterogeneity is warranted to arrive at estimands which support transportation of results to the population level or to centers where decisions are to be taken. In many meta-analyses, standard summary statistics reported in individual studies are presented in forest plots accompanied with a synthesis measure integrating result over the studies. With time-to-event outcomes, individual studies often summarize results as an (adjusted) hazard ratio (HR), representing the effect of different baseline exposure levels over time. This HR may be appropriately understood within one study where results are further accompanied by survival curves and an appreciation of the follow-up time along with descriptive statistics of baseline covariates in the observed exposure groups. One typically ignores, however, that in the absence of such context the meaning and hence the magnitude of HRs changes, leading such forest plots to combine apples with oranges in ways that are poorly understood.

Underlying assumptions must be made explicit for the interpretation of these pooled effect measures. If the studies are observational in kind, unequal distribution of potentially confounding factors may be natural and must be accounted for. To adjust for covariate differences between two exposure groups, case-mix corrected survival curves are used in observational studies with time-to-event data\cite{mackenzie_review_2012}. Also the censoring scheme could then be non-random (informative) and thus potentially influence the traditional outcome summaries\cite{howe_selection_2016}. Meta-regression can be used to assess the influence of covariates in meta-analyses of aggregated data\cite{thompson_how_2002}. However, information about influencing factors rarely is extractable from published reports in sufficient detail. To investigate confounding and informative censoring, one needs the Individual Patient Data (IPD). If there is sufficient information available regarding covariates in the IPD, the confounding can be determined and controlled. 
Pooling the HRs is a common way to perform a meta-analysis (MA) of time-to-event data\cite{parmar_extracting_1998}, but the HR as a summary for a time-to-event study is prone to different shortcomings as its interpretation is not straightforward\cite{hernan_hazards_2010} and depends on several study-specific characteristics\cite{jong_individual_2020}. Pooling of those study-specific HRs could, for instance, become uninformative if the study durations are different and the pooling ignores the baseline hazard which varies between studies. 
Applying a fixed-effects (FE) or random-effects (RE) model on those HR’s assumes a different underlying data structure\cite{jong_individual_2020, borenstein_basic_2010}. We are interested in the specific assumptions needed and the interpretations made to combine different studies into one pooled outcome statistic. 

In this paper, we propose a framework for MA that looks at the absolute survival differences between exposed and non-exposed groups as those are more intuitive and a better understandable model-invariant measure for clinicians. 
Our approach involves regression based on standardized survival curves and can be applied on both randomized and non-randomized studies with controllable and uncontrollable exposure. 

We illustrate this with an IPD MA of six studies  (\cite{koerber_influence_2014, yhim_prognostic_2011, balermpas_human_2017, meulendijks_hpv-negative_2015, mai_prognostic_2015, gilbert_p16ink4a_2013}) which assess the prognostic value of over-expression of the cellular protein p16\textsuperscript{INK4a} (p16) on survival among patients with anal squamous cell carcinoma. This is done as the question is posed whether p16 positive patients could receive a less aggressive treatment.
We applied the methodology on the individual data we received. The two main prognostic factors of interest in the original MA are: presence of DNA from oncogenic human papillomaviruses (HPV) and over-expression of p16, a surrogate marker of a transforming HPV infection, in the tumor\cite{obermueller_prognostic_2021}. 

\section{Methods}
To support future treatment decisions and enhance transportability of results, meta-analyses aim to identify sources of variation affecting outcomes across patients and studies. This capitalizes on within and between study heterogeneity but not without focusing on the question and hence imposing eligibility criteria. To learn about effects of point exposure on survival from (non-) randomized studies, we consider that different studies typically (1)~involve their own case mix in terms of baseline covariates, (2)~have different maximum study duration, affecting the questions we can address, (3)~are subject to their own censoring patterns, which may differ by design, by the nature of their population or by exposure (effects), (4)~focus on different primary end points and (5)~suffer from various levels of missing data. 
Each of the above impacts on the questions that can be addressed, on the possible biases and efficiency problems to tackle, both at the analysis and reporting stage.

\subsubsection*{Beyond hazard ratios in meta-analyses}

To pool HRs, as summary measures for the exposure effect in each study, one implicitly assumes these to remain constant over time in each study. Hernán explains \cite{hernan_hazards_2010} why the usage of these HRs could be problematic. The time-constant HR estimate is a time weighted average of period specific HR’s. It thus changes as the follow-up times change  unless the true underlying HRs are time-fixed. 
In MA of studies with different average or maximum follow-up duration, it may be more meaningful to pool time-specific HRs over bounded time periods\cite{hernan_hazards_2010}. This would limit the use of methods which try to pool conditional survival probabilities in order to reconstruct the pooled survival probability as suggested by Takagi\cite{takagi_meta-analysis_2014}. 

The HR may thus depend on the chosen or observed maximum study time. When pooling the HR for a MA, we could choose to neglect longer observation time of certain studies and limit the period of interest to the shortest study in order to average over parameters with similar population level meaning. However limiting studies in time would alter the used HR of that study compared with the original reported study specific HR in the original study. We would also loose the possibility to look beyond this early stopping point and could no longer detect certain longer term changes. Limiting the MA to the shortest study time could encourage the exclusion of studies with short study durations, which leads to missed studies.

Wei et al.\cite{wei_meta-analysis_2015} used the difference in the restricted mean survival time to allow MA of time-to-event outcomes with non-proportional hazards.

Different methods exist to control for confounding. One-stage or two-stage approaches are used to perform a MA. In a two-stage approach, the estimator is first calculated for each study individually and subsequently those estimators are pooled. This is also done in meta-analyses of aggregated data where the summary statistic is directly extracted from published papers and where the IPD are not available. The heterogeneity is determined also in two steps, whereas in the one-stage approach within and between study heterogeneity are calculated simultaneously\cite{schmid_handbook_2020}.

Vo et al.\cite{vo_novel_2019} proposed a framework to reduce this case-mix heterogeneity appearing in randomized controlled trials and infer treatment effect in a well defined case-mix population. They propose to use a specific study population as the reference population but suggest clinicians to use this approach to provide effect measure estimates standardized to the case-mix distribution of an external reference population. 

\subsubsection*{Descriptive statistics and unavoidable standardization in Kaplan-Meier curves}

Even at the descriptive stage, survival outcomes may need work -due to censoring- before meaningful summary statistics can be given. 
The maximum and observed follow-up time, $\tau$, are typically known or fully observed in each study s and exposure e group, as are the baseline covariates ($\mathbf{Z}$) that may be prognostic for survival. If the distribution of the latter changes over time, we must condition on them to avoid informative censoring due to end of study, i.e. due to administrative censoring. There may be other mechanisms too that render some prognostic baseline covariates ($\mathbf{Z}_c$) predictive of the censoring time.

Even non-parametric Kaplan-Meier (KM) curves per study and exposure group are then unable to the survival experience of each (s,e) group: they suffer from systematic bias due to informative censoring. Conditional on the set of baseline covariates $\mathbf{Z}_c,$ however, censoring becomes  non-informative provided : $C \cip D| s, E, \mathbf{Z}_c$, with C representing time of censoring and D time of event. As a result the standard (or as needed extended) Cox model can be fit conditional on $s$, $E$, $\mathbf{Z}_c$ under the usual non-informative censoring assumption. The resulting  hazards $\lambda (t|s,e, \mathbf{Z}_c)$ can then be turned into conditional survival curves $S(t|s,e, \mathbf{Z}_c)$  which are correct. Upon averaging these over the (s,e)-specific covariate distributions of $\mathbf{Z}_c$ we obtain what the KM curve is generally expected to measure. 

Well chosen case-mix corrected (or adjusted) survival curves are known methods in the literature\cite{hernan_hazards_2010, mackenzie_review_2012} that can not only avoid confounding due to censoring but also account for covariance imbalance and transportability of obtained results. Last but not least important for the MA: these survival curve estimators aim to show interpretable differences.

\section{General framework}

Clinical trials evaluating the same treatment over different centers naturally work with somewhat different patient populations. Part of the differential case-mix typically seen in the distributions of observed baseline covariates per center and the remaining differences are caught by the study-specific baseline hazards. In the randomized setting, measured and unmeasured confounders are exchangeable between treatment arms within study, but the study-specific treatment effect likely varies between centers. This happens either by variation in the administered treatment and/or differential action of the exact same treatment over the various environments.

Different estimands can be targeted before pooling center specific effect measures. For fair comparison between centers, one may average in each center $s$ the $\mathbf{Z}$-conditional treatment effect, $S_s(t; \mathbf{Z},E=1) - S_s(t; \mathbf{Z},E=0)$ over a common patient mix of measured prognostic factors, e.g. the joint $\mathbf{Z}$  distribution across all $k$ centers.
Additionally one could fix the baseline hazard, representing a center-specific patient mix of unmeasured prognostic factors or involve every patient of the combined population with their own measured covariates $\mathbf{Z}_i$ and the baseline hazard of their own center $\lambda_{0,s}$.  


In this paper we evaluate the effect of a point exposure $E$ at baseline, based on k different studies (s = 1, 2 ,.., k) with time-to-event data.  We start from a common set of centered baseline covariates, $\mathbf{Z}$, and present a fully general `proportional' hazards (PH) model. In equation \eqref{eq:model}, study-specific covariate effect functions $\bbeta_s(t)$,  exposure effect function $\psi_s(t)$ and interaction effect function $\bgamma_s(t)$ act on the study-specific baseline hazard $\lambda_{0,s}(t)$. The latter thus represents survival for the reference exposure level $E_i=0$ across center $s$. These effects could be time-dependent, while $g$ and $f$ may be nonlinear functions of $(\mathbf{Z}_i, E_i)$:

\begin{equation} 
\lambda_i(t|s_i,\mathbf{Z}_i,E_i) = \lambda_{0,s_i}(t) e^{\left( \bbeta_{s(i)}(t)*f(\mathbf{Z}_i)+\psi_{s(i)}(t)*E_i+\bgamma_{s(i)}(t)*g(\mathbf{Z}_i,E_i) \right) }
\label{eq:model}
\end{equation}

In case we restrain ourselves to those without an exposure (E=0) in the RCT setting and the covariates are centered, the study specific baseline hazards capture the true differences between studies centers, which could correspond with intrinsic differences of the study populations or standard care practices of the study center. For practical purposes, we simplify the situation to only allow an additive exposure effect and assume time constant covariates and exposure effects. This simplification could represent a baseline situation in an RCT with the addition of a single exposure impact. Table \ref{tab:overview_sources} displays the different sources of variation relevant to (non-)randomized studies.

Where the true HR is time-varying, the working PH Cox model involves a weight averaged summary over time of $\psi(t)$ with weights dependent not only on the distribution of the event time but also on the distribution of the censoring time, with varying study duration these hence target different estimands\cite{xu_estimating_2000}.

\subsection{Study representation}

We propose a set of questions and explicit statements which help researchers to choose the most appropriate estimand and estimator and facilitate its interpretation and meta-analytic pooling.

With standardization, we address a number of `what if' questions, which may or may not have causal interpretation, but seek to put treatment effects summarized in different studies on a more equal footing e.g.: What would the overall survival be at three years if patients from center x had all been given basic care at center x in an RCT? Could we describe the treatment effect for those patients as the difference in overall survival for all those patients assuming they all have been given the treatment? If that treatment effect depends on the covariate distribution at center x, should we define a common covariate distribution among all centers in order to have meaningful comparisons? 
What treatment effect would patients of center x with actionable exposure or treatment given in an observational setting experience, if they were patients in center y and what is the impact of the observed different covariate distribution of that patient population? What if we are concerned with the non-exposed population only and wish to evaluate their expected survival experience under the non-exposed versus the exposed conditions in each of the centers?
In case patients are exposed to a non-actionable exposure, one could be interested in what would the overall survival be for the group of patients without the exposure if they all got the basic care in center y and how would this relate to the overall survival for exposed patients if they would all receive the treatment.

What if the different centers are evaluated over a similar follow-up time, have similar distribution of measured covariates, additionally similar unmeasured covariates reflected in a common baseline hazard (e.g. representing the non-exposed condition for patients with average characteristics)?


\subsection{Adjusted survival curves}

We represent the individual study s, with sample size $n_s$, by regressing outcomes over a certain set of covariates (\textbf{Z}), which is referred to as the marginal approach: $\frac{1}{n_s}\sum_{i=1}^{n_s}S_s(t|\mathbf{Z_i})$.
Or we could use a profile specific approach: $S_s(t|\mathbf{Z_i}=\mathbf{z})$.
In the latter we would have to choose one specific set (\textbf{z}) of covariates which represent a(n) (fictional) observation in the study.

We first focus on the marginal approach. Depending on the estimand of interest, choices of the distributions over which we average (\textbf{Z}, $\lambda$, $E$, $\bbeta$ and $\psi$) must be made in order to obtain a certain adjusted survival curve. This survival curve could be the final statistic for the MA to pool or an intermediate stage, which requires further manipulation to obtain the contrast of interest, for example risk difference at certain time point or the return to a marginal HR.

We will present five different approaches to standardize the outcomes. Each of them requires different assumptions and has different interpretations.

\subsubsection*{Marginal approach 0}

One would naturally start by describing the observed survival experience in exposure groups per study, revealing heterogeneity. It is a starting point from which to explore the role of specific variance components. When baseline covariates ($\mathbf{Z_c}$) describe censoring time besides survival time, hazards will suffer from explainable informative censoring. Without accounting for ($\mathbf{Z_c}$) the hazards do not just reflect the survival experience.
Should observed time-varying covariates explain both censoring and the hazard of interest, inverse probability of censoring weighting is a way forward\cite{robins_correcting_2000}.

Descriptive statistics for right censored survival outcomes typically show KM curves to give a sense of the marginal outcome distribution (per study and exposure group) over the study duration. These curves will be biased if some baseline covariates affect both survival time and censoring time. When this happens, marginalized survival curves may be obtained as standardized covariate specific survival curves, averaging over the covariate distribution of the study-specific exposure group $s_e$. As described above, this can be done, for instance, using a PH model. 
With $n_{s_e}$ the size of observed  study population  s  exposed to level e, the study and exposure specific survival equals:

\begin{equation} 
\label{eq:marg_approach_0}
S_{s | \ctb{s_e}}(t)=
\frac{1}{\ctb{n_{s_e}}}
\sum_{\forall i \in \ctb{s_e}} S(t|\ctb{\mathbf{Z}_i},E_i,\lambda_{\ctb s}, \bbeta_{\ctb s}, \psi_{s})
\end{equation}

These survival curves reflect the patient mix of study specific exposure groups. Contrasts of such curves for different levels of exposure are potentially confounded, but such a curve is designed to reveal the realistic survival experience in each concerned subpopulation. To evaluate how the exposed group in study s might have fared without the exposure (treatment effect among the treated) we would match $S_{s|\ctb{s_e}}(t)$ with a counterfactual unexposed counterpart. Alternatively, we can perform indirect standardization, for instance taking the full study s population as the target patient mix as we describe next.

After this regression we still expect to have within and between study differences for the survival based on different covariate distribution within and between the exposure groups and studies. We are also able to do this without making many assumptions as we regress each exposure group within the study over the own exposure specific covariate distribution.

\subsubsection*{Marginal approach A}

In the next approach we reduce within study variation by addressing the case-mix heterogeneity of the exposure groups. We will standardize the outcomes at set exposure level e over the study s specific entire distribution of covariates \textbf{Z}. For set exposure e:

\begin{equation} 
\label{eq:marg_approach_A}
S_{s,e | \ctb{s}}(t)=
\frac{1}{{\ctb{n_s}}}
\sum_{\forall i \in \ctb{s}} S(t|{\mathbf{Z}_i},e,\lambda_{s}, \bbeta_{s}, \psi_{s})
\end{equation}

The exposure status $e$ is transported to the whole population of study s. For this to make sense, involved covariates $\mathbf{z}$  must  have non-zero probability of being exposed at exposure levels considered.  We would interpret $S(t|{\mathbf{z}_i},e,\lambda_{s}, \bbeta_{s}, \psi_{s})$ as the outcome distribution that would reign if the patient had exposure e, as the counterfactual outcome.
Provided there are no unmeasured confounders (after accounting for $\mathbf{Z}$), we now can compare the within study survival difference at time t for two set levels $e_1$ and $e_2$ as the expected difference in survival chance at time t, if study population s were subject to $e_1$ rather than $e_2$. When this absolute risk difference at one year is 5\%, say, we can expect to have $0.05*n_s$ more surviving patient at one year if exposure level $e_1$ reigned rather than $e_2$. Comparing  this measure between studies will however still embrace  case mix heterogeneity and differences due to the baseline hazards in addition to genuine exposure effect differences between studies. 


\subsubsection*{Marginal approach B}

The third approach incorporates the  \textbf{Z} distribution of all studies combined, study population $\Omega.$ We will standardize the exposure e outcomes at study baseline hazard $\lambda_s$ over the full combined distribution of covariates \textbf{Z} using the study s specific covariate effects $\bbeta_s$. For set exposure e we define the survival for study s given the study specific baseline hazard while regressing over whole population: 

\begin{equation} 
\label{eq:marg_approach_B}
S_{\ctb{s}, e ,\lambda_{\ctb{s}} | \ctg{\Omega}} (t) =
\frac{1}{\ctg{n_{\Omega}}}
\sum_{\forall i \in \ctg{\Omega}} S(t|\ctg{\mathbf{Z}_i},e,\lambda_{\ctb s}, \bbeta_{\ctb s}, \psi_{\ctb{s}})
\end{equation}

Here we examine what would happen if baseline conditions (and hence baseline hazard) and covariate effects $(\bbeta_{\ctb s})$ of study $s$ would continue to operate under various possible exposure levels $e$, combined with the observed covariate $\ctg{\mathbf{Z}_i}$ distribution of the full study population.  

It emulates outcomes upon  transporting all patients to study s with set exposure level $e$, assuming these groups would solely differ in the distribution of \textbf{Z}, and hence not in any unmeasured predictors $U$ (Such as gene pool or socio-economic factors, environmental conditions, etc.) which are bound to affect the baseline hazard and possibly covariate effects $\bbeta$. To make practical sense, the covariates $\mathbf{Z}$ should also have the exact same definition and representation in each study. As between study heterogeneity is to be expected, also in randomized trials, baseline hazards $\lambda$ are unlikely to be  equal over all studies. The thus standardized survival curve becomes a more theoretical concept, allowing to compare exposure effects between studies on similar grounds. One should think about extending the s-population to cover the full Z-distribution among its patients, without transporting actual patients or exposures from one center to another.   

We propose to use the whole population $\Omega$, and will use the counterfactual outcome in which we do not only set the exposure, as done in approach A, but we will use the counterfactual outcomes for a patient $i$ set with study specific effects: the baseline hazard, the potential exposure effect and the covariate effect. We need the positivity assumption, chance to be included in different study and to experience both exposures in those different study groups, to be valid.

\subsubsection*{Marginal approach C}

Our next approach relaxes some of the strong assumptions made in approach B. We standardize the exposure e outcomes for study s by regressing over the full combined distribution of covariates \textbf{Z} using the study specific covariate effects $\bbeta_{s(i)}$ and the study specific baseline hazard $\lambda_{s(i)}$ of the individual:

\begin{equation} 
\label{eq:marg_approach_C}
S_{\ctb{s},e | \ctg{\Omega}} (t) =
\frac{1}{\ctg{n_{\Omega}}}
\sum_{\forall i \in \ctg{\Omega}} S(t|\ctg{\mathbf{Z}_i},e,\lambda_{\ctg{s(i)}}, \bbeta_{\ctg{s(i)}}, \psi_{\ctb{s}})
\end{equation}

We now transport exposure effect $\psi_s$ of study s to act on all patients i with their study s(i) specific  baseline hazards and covariate effects.  We then average over the resulting potential outcomes. When exposure e is set to 0 for all,  this yields the potentially unexposed  population average outcome, our reference value $S_{\ctb{s},e=0 | \ctg{\Omega}}(t)$. Note how it is independent of the  study s, whose effect we examine. If $\psi_s$ primarily reflects the impact of a treatment/exposure level e as it would be administered (with surrounding care) in center s, we may thus wish to evaluate what would happen if all patients were subject to this type of treatment/exposure, but otherwise carried their own gene pool and center specific covariate outcome association.           
Unlike in approach B we now consider transporting all patients with their governing  baseline hazard and covariate effects  to the setting of each study $s$ with its exposure effect in turn. They arrive there with both their measured  \textbf{Z} and unmeasured \textbf{U} prognostic factors. This is realistic as long as exposure effect $\psi_s$  does not interact with baseline hazard and/or covariate effects. The model(s) could however easily be extended to allow for a $\mathbf{Z}$ by $e$ interaction effect that is study specific, $\psi'_{Z,s}$.
We still need a positive probability that patients with covariates \textbf{Z} could have the exposure level e and would be part of study s, which is also needed in approach B.

\subsubsection*{Marginal approach D}

Finally we propose approach D which regresses the survival outcomes over the covariate distribution of the patients with a specific exposure, $\mathbf{Z}_{E=j}$, using the baseline hazard of the study $s$:

\begin{equation}
\label{eq:approach_D}
\begin{split}
S_{\ctb{s}, e ,\lambda_{\ctb{s}} | \ctg{\Omega_{E_i=j}}} (t) & = \frac{1}{\ctg{n_{\Omega_{E_i=j}}}} \sum_{\forall i \in \ctg{\Omega_{E_i=j}}} S(t|\ctg{\mathbf{Z}_i},e,\lambda_{\ctb s}, \bbeta_{\ctb s}, \psi_{\ctb{s}}) \\
\end{split}
\end{equation}

This would balance the exposure groups over the different studies and within each study with respect to the covariate distribution. These survival curves would show the variability between and within studies and would no longer depend on differences in measured covariates. This approach helps to understand the variability of e.g. the treatment effect among treated or the exposure effect among unexposed.

We need positivity for the chosen exposure group to be included into the exposure specific subgroups of study s. 


\subsubsection*{Conditional approaches}

The regression method could be avoided by taking a conditional approach. We could define a relevant profile and perform the MA conditional on this choice. The population within and between studies could be very heterogeneous, determining this \textbf{z} is thus not trivial. As in the marginalizing approach, we can choose a profile per exposure subgroup, per study or per exposure and study subgroup. If we picked one profile for all exposures and studies, we could easily compare between the different exposure levels and studies, as the outcome would no longer be dependent on the measured confounding represented in the profile.

If we choose an “average” person, represented by a summary measure of the characteristics, the interpretation of such a person could become meaningless, e.g. the fraction of gender has no meaning in this context. An illustrable "average" person could be considered as the observation with the median of the prognostic scores which would indeed represent a real person. We could use the 10\% and 90\% quantiles of the prognostic score to quantify the variability of that population. The profile we would choose is best clearly situated within the study population, thus a non zero-chance of having that profile in a particular subgroup. This is applicable to only one profile, however, when using the regression methods, we need to be sure that they all are representative. 

\subsection{Estimands of interest}

\subsubsection*{Exposure contrast}


We need to define further the estimator of interest. The effects could be described on the hazard scale or the translation could be made to the risk or odds scale. Further, we can distinguish time specific measures, at prespecified and clinically relevant time points, and time averaged measures.

At the risk scale, we could (i) keep absolute survival chances over time for the different exposures, (ii) express exposure contrasts relatively in ratio’s or (iii) in differences of distinct exposures.  
Absolute survival chances give an understandable result per study, not neglecting the meaning of the chosen standardization, allowing a clinician to judge the relevance of potential differences and the course in time. 
In a setting which involves a selectable treatment, the survival difference at time $t$ relates to the number of patients needed to be treated to save a live. This holds true, if we have an approach in which the survival curves for each treatment represent a similar population. In the setting of selectable exposures we can have the analogue. If the exposure is non-controllable or stochastic, this clear analogy is lost.
The ratio of survival chances could be interpreted as relative risks. 

A time specific measure on the hazard scale could be possible, absolute or relative, but the interpretation of conditional rates is not very intuitive. The difference in hazards between the two exposure groups, for instance parameterized in an additive hazards model, is not discussed in this paper. Both would still require to be evaluated at a time of interest, preferably with clinical relevance. 
Typically one uses the obtained conditional HR of a PH model in the MA in a standard two-stage method. A benefit is the usage of published results from literature. 
Daniel and colleagues proposed a procedure for estimating the marginal HR for an exposure effect using standardized survival functions after fitting a PH model conditional on covariates \textbf{Z} using simulations\cite{daniel_making_2020}. If one uses an averaged measure, which represents the overall survival curve, the maximal time of the survival curve considered still has to be chosen. Especially if one allows for a piecewise constant exposure effect.

\subsubsection*{Pooling and forest plots}

Questions which remain are which model we will use for the pooling: fixed or random effects and the accompanied weighting schemes.

In the context of standardized mean differences Marin-Martinez and Sanchez-Meca recommend to use Hedges and Vevea’s estimator instead of Hunter and Schmidt, based on the individual study sample sizes\cite{marin-martinez_weighting_2010}. Shuster argues that for a patient level effect size in MA of different trials, the total sample size of each study should be used in the weighting of the study effect sizes\cite{shuster_empirical_2010}.

\section{Results}

\subsection{Motivating example}

\subsubsection*{Descriptives and censoring}

We applied our methods to the IPD from six independent studies. These observational studies gathered data about the overall survival time of patients with anal squamous cell carcinoma treated with radio-chemotherapy, radio-chemotherapy and surgery or radiotherapy alone. The study populations were rather heterogeneous with respect to prevalence of p16 and HPV DNA positivity, gender, age, staging and other covariates. The maximum duration of follow-up varied between 72.5 to 325.4 months. (See table \ref{tab:overview_covars} for a descriptive overview.)

\begin{figure}[h]
\centering
\includegraphics[width=0.75\textwidth]{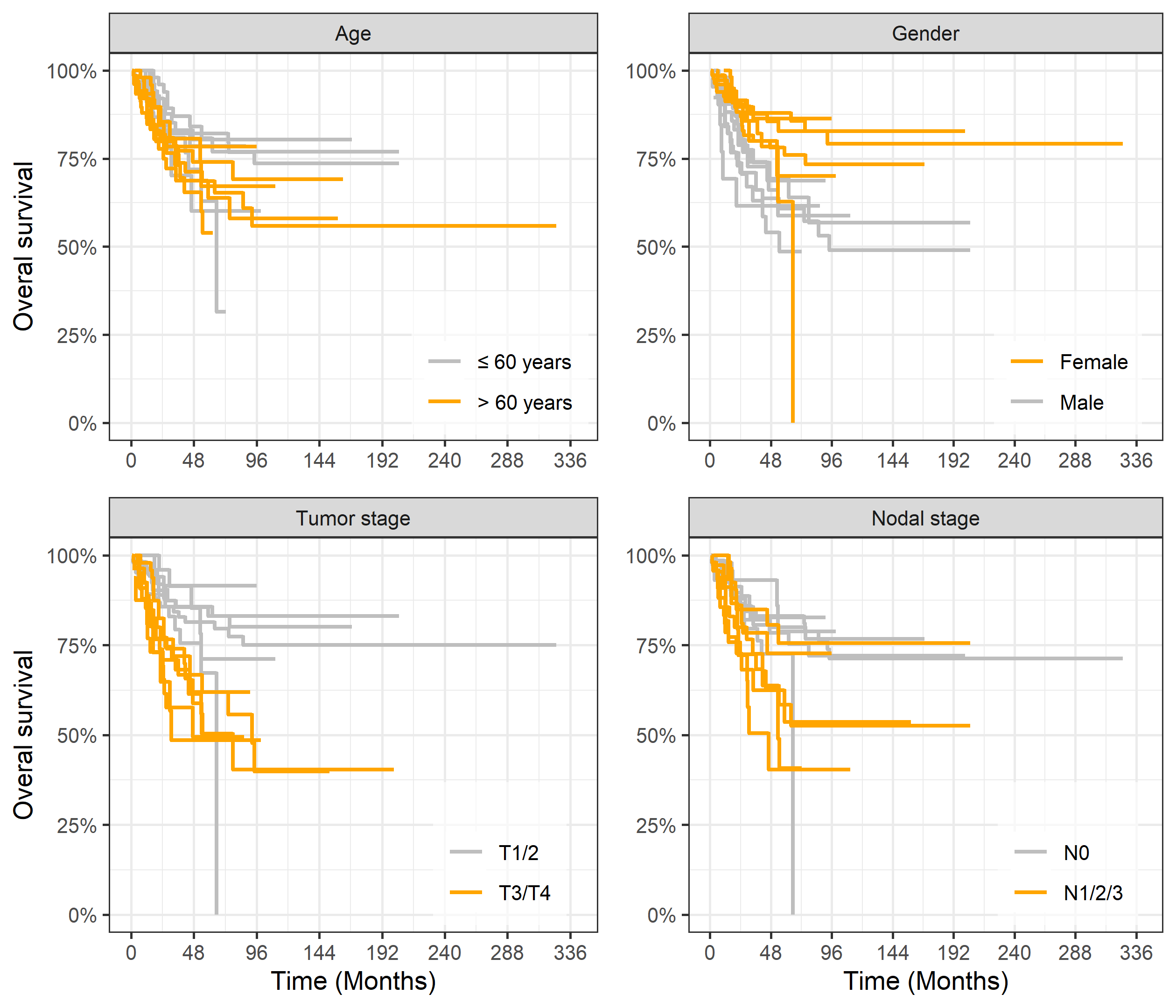}
\caption{Study specific Kaplan-Meier curves stratified subsequently per age category, gender, tumor stage group and nodal stage group for the 615 complete case patients. }
\label{fig:KMperCovar}
\end{figure}

Overall survival according to age (dichotomized at $>60$ years), gender, T-stage or N-stage was calculated per study, see figure \ref{fig:KMperCovar}. It shows the potential confounding and the  differences in maximum follow-up time between the studies and subgroups. Younger patients and patients with less advanced stages of tumor show higher overall survival.

\subsubsection*{Proportional hazards model}

A PH Cox model was fitted that was stratified by study to assess informative censoring. It was adjusted for one continuous variable, age, and three categorical variables: gender, tumor stage (T) group and nodal stage (N) group. We assumed a common effect for age, gender, T- and N-stage over all the studies, while we allowed the effect of the exposure variable p16 to differ among studies. Gender, tumor stage (grouped T3/T4 versus T1/T2) and nodal stage (grouped N1-3 vs N0) were entered in the model as binary variables, while age was considered a continuous variable. We did not allow for interaction terms between the exposure and the covariates.
Age and T-stage showed to be rather informative with HR of 1.083 [95\%CI 0.996-1.177] for a ten year increase in age and a HR of 1.230 [95\%CI 0.992-1.526] for tumor stages T3/T4 vs stage T1/T2. Older patients and patients with more advanced tumor stages tended to be censored sooner.

The maximum observed follow-up time for p16 negative patients varied between 33.6 and 325.4 months, while for p16 positive patients this varied between 72.5 and 201.0 months. 

\subsubsection*{Flexible parametric PH model}

We applied a flexible parametric PH model using the rstpm2 package (version 1.5.2)\cite{liu_parametric_2018} in R. The study stratified baseline hazard is modelled with a natural cubic spline function parameterized by $\bzeta_s$ and fitted on the logarithm of time scale. The confounding covariates (age, gender, T-stage, N-stage), the exposure (p16 positivity) and the interaction between exposure and study are considered proportional effects to the hazard function. With $\mathbf{Z'}^T = [\mathbf{Z}^T, E, s, E*s]$:

\begin{equation} 
\begin{split}
\log(-\log S(t|\mathbf{Z},E,s)) & = \log H(t|\mathbf{Z},E,s) \\
 & = \bbeta'^T\mathbf{Z'} + \log H_s(t|\bzeta_s)    
\end{split}
\end{equation}

The model assumes a common effect for the covariates \textbf{Z}, because of limited true events. This is a strong assumption as one could allow study specific \bbeta. The study specific baseline hazard is a combination of two cubic splines which goes from the minimum logarithm of the true event times until the median true event times, and from the median true event time to the maximum true event time. These times are defined equally for all the studies. The splines are defined as linear functions before and after the boundary knots. By using one internal knot the suggestion of Royston and Parmar was followed\cite{royston_flexible_2002}. In our example the model was restricted to not have negative hazards, not automatically fulfilled by the software. 

\begin{figure}[h]
\centering
\includegraphics[width=0.8\textwidth]{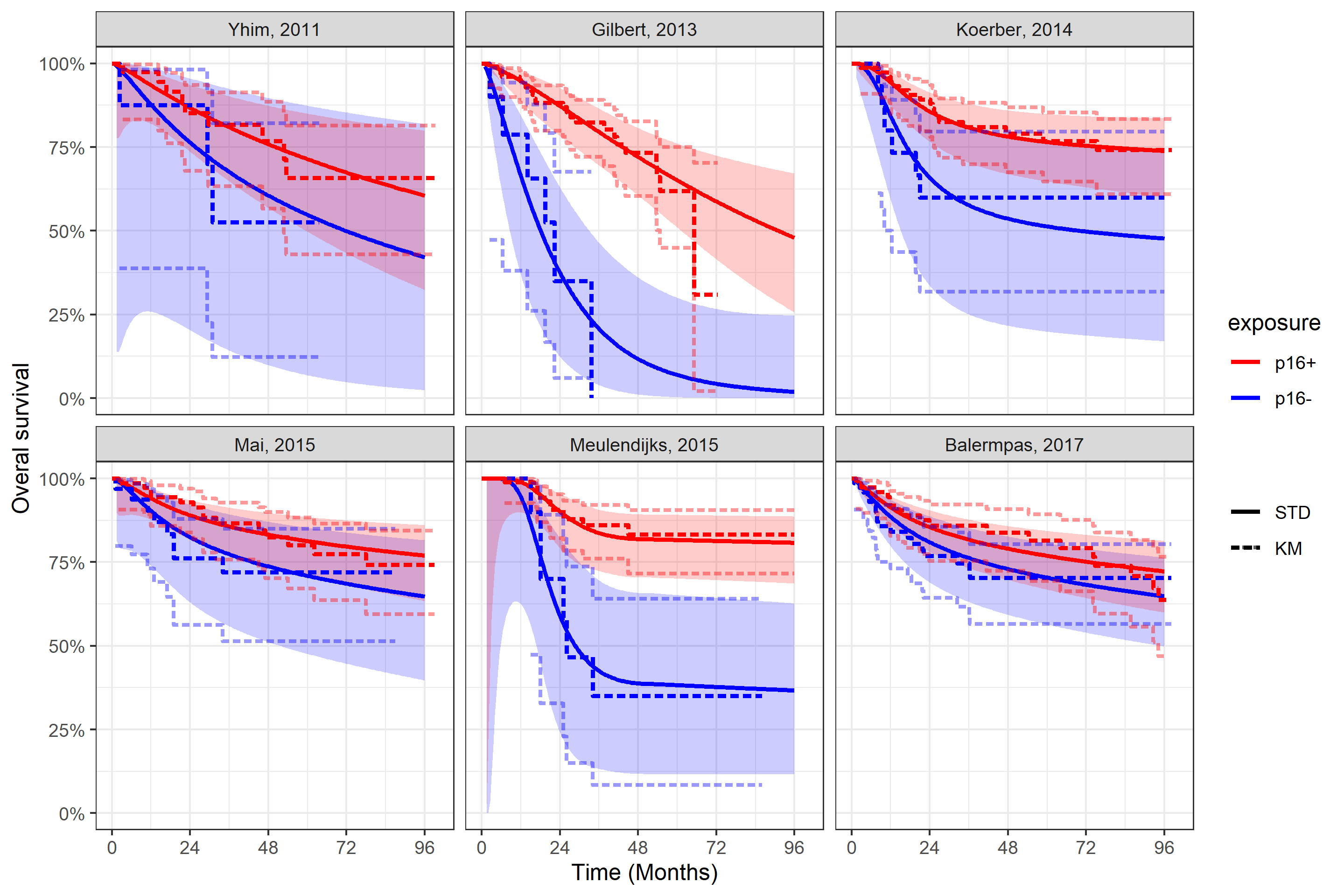}
\caption{Kaplan-Meier (KM) and standardized survival curves (STD) per study and exposure (dotted lines, respectively shaded area, as the 95\% pointwise confidence interval). Standardized curves, using approach 0, are derived from a flexible parametric proportional hazards model stratified per study and fitted to the complete case population. Standard errors are obtained using the jackknife method, with hazard estimates restricted to be positive.}
\label{fig:KM_STFLEX_perstudy}
\end{figure}

In figure \ref{fig:KM_STFLEX_perstudy} the naive KM curves per study are shown for a certain time period. The adjusted survival curve with approach 0 (equation \ref{eq:marg_approach_0}) was also calculated using the flexible parametric PH model and shown together. This standardized survival curve does no longer suffer from informative censoring. In the case study, the unexposed group, anal cancer patients with a p16 negative tumor, have in all the studies a smaller overall survival chance than the exposed group.

\begin{figure}[h]
\centering
\includegraphics[width=0.99\textwidth]{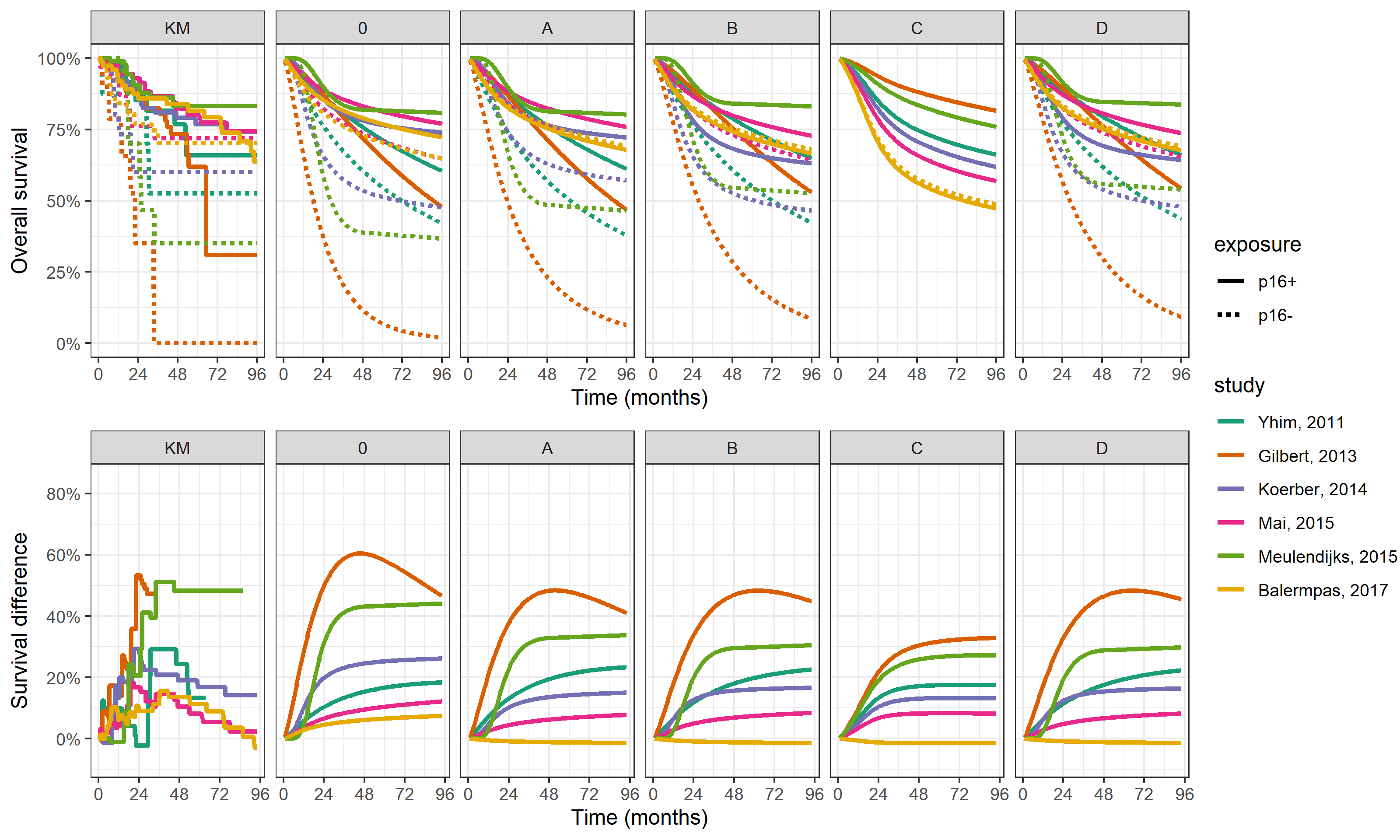}
\caption{The Kaplan-Meier (KM) curves and the different standardization approaches using the flexible proportional hazards model with the respectively derived survival differences per study.}
\label{fig:ST_diff_ABS}
\end{figure}

\subsubsection*{Estimand of interest}

p16 positivity is not an attributable intervention and the clear interpretation of applying approach B or C is somehow lost. 

For illustrative purposes we present here all the standardized survival curve approaches, pointing to the differences between them. We focus on the standardized survival differences at 12, 24 and 36 months and the marginal HR at 36 months and 60 months. 

We first used all the available data and calculated the marginal HR for the standardized survival curves (using all the different approaches) for time periods $[0-36 \text{ months}]$, $[0-60 \text{ months}]$ and $[0-94 \text{ months}]$, see figure \ref{fig:forest_margHR_uncensored}. We then repeated this but before calculating the PH flexible model we artificially censored the studies at 60 months, see figure \ref{fig:forest_margHR_censored}. For the absolute scale we looked at the 1, 2 and 3 year risk difference, also here first with the model of the full data (Figure \ref{fig:forest_survdif_uncensored}) and with the model applied to the artificially administrative censored data (Figure \ref{fig:forest_survdif_censored}). 

\subsubsection*{Standardized survival curves}

Figure \ref{fig:ST_diff_ABS} shows the KM curves and the different standardized survival curves for every approach and study. The comparison of the risk difference obtained with the KM curve and the approaches using a flexible parametric model shows the benefit of this more smooth behavior. If we would have used a Cox model, this step function with steep changes in survival risk would still have appeared. But care has to be taken as soon as parametric modeling allows for extrapolation beyond what is seen in the KM curves. In table \ref{tab:overview_covars} the total study duration (and time to last death) is given per exposure group, restricting survival measures to 36 months avoids making assumptions of the extrapolation.

\subsection{Meta-analysis}

For the pooling we applied a random effects model with the DerSimonian-Laird estimator\cite{dersimonian_meta-analysis_1986} and the Knapp and Hartung adjustment\cite{inthout_hartung-knapp-sidik-jonkman_2014} using the metafor package (version 3.0-2)\cite{metafor} in R. 
The prediction intervals for the pooled RE estimate are calculated according to Riley\cite{riley_interpretation_2011}. 

The time specific survival differences per approach (Figure \ref{fig:ST_diff_ABS}) visually show minor differences between approach A, B and D in our case data set. For approach C, the survival difference functions per study do not cross each other. Note that not all studies had a follow-up of 96 months and extrapolation occurs. We applied the MA of the survival difference at multiple time points, i.e. 12, 24 and 36 months, the forest plots are shown in figure \ref{fig:forest_survdif_uncensored}. We additionally applied artificial administrative censoring on the data before fitting the model and applying the standardization, shown in figure \ref{fig:forest_survdif_censored}. For the first we obtained very similar pooled survival differences at 36 months, where only approach 0 and C were significant beneficial, in the latter the pooled survival difference at 36 months showed a small increase compared with the non-artificial censored data and all approaches showed a significant beneficial effect. 

If the data came from an RCT where the exposure would have been the treatment, we could estimate the pooled treatment effect at 36 months as the survival difference at that time using approach C, 0.15 [95\% CI 0.03-0.27], or as a pooled marginal HR for the standardized survival curves until 36 months, 0.79 [95\% CI 0.66-0.94], both showing a significant beneficial treatment effect. 

In table \ref{tab:overview_HRs} an overview of the HRs from the different studies is given. The crude HR without any correction for confounding, the conditional HRs obtained from the model and the marginal HRs after applying approaches A and C with their considered time-periods. Depending on the approach, the 95\% CI of the pooled estimate does or does not include unity. Choosing a priori an approach and stating which assumptions one could make is of importance. In our case study the pooled effect based on the crude HR overestimates the exposure effect compared with the marginal HRs.

\section{Conclusion}

Reporting time-constant HRs in forest plots is common practice in MA of treatment effect on survival. When not all HRs are truly time-constant and study duration or censoring patterns differ between studies, such plots may reveal between study differences that are driven by observation characteristics (censoring) rather than true survival experience. It happens even more when non-randomized studies suffer from covariate imbalance between exposure groups at baseline. 

In this paper, we have shown how standardized survival curves may avoid spurious heterogeneity allowing forest plots to focus on relevant survival contrasts within and between studies. Typical standardization efforts are conducted within studies, to avoid informative censoring and to evaluate outcome in exposure groups which are balanced in terms of baseline covariates. For the MA we may additionally wish to seek balance between studies. This raises new questions about the relevant reference distribution and the role of the baseline hazard in such a setting. 

We developed different ways of balancing (un)measured covariates within and between studies, leading to distinct but clear interpretation of contrasts between survival curves stripped from unwanted sources of variation.
We discussed various options and show how the corresponding survival curves and their contrasts differ in meaning and magnitude. 

Approach 0 would give an unbiased outcome per study if censoring was informative but explainable by covariates. Approach A yields treatment contrasts per study while fully embracing population heterogeneity between studies while approach B and C do correct for this using a common distribution of measured covariates on top of either a study-specific or shared baseline hazard,    each with its own interpretation and assumptions.

Using observational data, one should avoid the naive KM and go beyond approach 0 if confounding is present. For cases with actionable treatments, approaches A, B and C are designed to emulate an RCT for different eligible populations. Approach D presents the differences caused by exposure for the study subpopulations defined by their observed exposure level.

If one does not find it plausible or helpful to transport observed effects to other study centers, our method still provides a framework that facilitates decomposition of the variance seen within and between studies and yields corresponding insights.

\section{Discussion}

Although our development on MA for IPD was motivated by an evaluation of exposure in observational studies, many of the issues raised and implemented are just as relevant for MA of randomized trials. We consider the many criticisms of reporting survival studies primarily in terms of HRs since it limits clinical interpretation and weakens the evidence base on which decisions are relying.
At its worst comparing HRs across centers can be misleading since they relate to more than 'just' the survival experience. We propose a framework for MA reporting that can start from the standard PH modeling approach (or an adaptation thereof) but derives standardized survival curves which bring a more transparent and interpretable report on variation between studies in exposure related survival averaged over a common patient mix and potential confounders of censoring. 

In clinical papers study population characteristics are typically presented  in the mandatory summary table 1. Baseline characteristics are important to understand the different populations which are contained in the MA. The absolute survival at a clinically relevant time point, calculated using approach 0 for the unexposed group remains of interest as it will inform readers on the survival experience in the observed exposure groups, the heterogeneity in survival between studies under control conditions. In RCTs, this non-exposed or non-treated survival may reflect a mix of other treatments. For observational studies, the between study survival differences are similarly impacted by differences in baseline characteristics. The interpretation would require similar caution as for the between exposure groups differences.

The IPD of the different studies are required to use our framework. However in general these IPD are not easily obtained. We would recommend current publishing authors to provide KM curves jointly stratified for all major covariates with corresponding risk tables, to allow for the derivation of standardized survival curves\cite{wei_reconstructing_2017}.

To allow for adjusted case-mix over studies, they must register the same set of covariates with the same definition for those covariates. We did not handle cases with missing covariates (or exposure status) in which, potentially, one could use multiple imputation. 

We mentioned the importance of maximal follow up when averaging time dependent HRs. One could further assess the impact of restricting the MA to studies with a minimal study duration and the consequences of studies with poor follow-up.

We showed multiple approaches to standardize survival curves, and recommend to carefully think about which estimand fits best and what assumptions they involve. Important are the aspects of transportability: We used the combined population of all studies in approaches B, C and partly D, which require stronger assumptions than approach A where we used per study the study specific population. 

Expressing survival contrasts as survival differences ease interpretation and relevance. Estimands relevant from a public health perspective in observational studies, contrast the current population with all treated populations under clear assumptions. Next, we advocate to show the standardized survival curves per exposure group over a clinically relevant time period. If one uses a parametric approach and uses the advantage of extrapolation, this must be explicitly mentioned. 

We did not handle some important aspects when some assumptions are violated, such as unbalanced group, for which the positivity assumption is no longer valid, study specific measurement errors or problems with classification of exposures or covariates, external -unmeasured-factors which influence only certain studies. Although we simplified our model and did not allow for interaction effects of a confounder with the exposure, the usage of our adjusted survival curves remains.  









\section*{Acknowledgements}
Mark Clements (discussion on applying rstpm2 R package) and Liliana Belgioia (Health Science Department (DISSAL), University of Genoa, Department of Radiation Oncology, IRCCS San Martino
Hospital, Genoa, Italy), Duncan C. Gilbert (Sussex Cancer Centre, Royal Sussex County Hospital, Brighton, UK), Stefan A. Koerber (Department of Radiation Oncology, Heidelberg University Hospital, Heidelberg, Germany), Sabine Mai (Department of Radiation Oncology, University Medical Center Mannheim, University of Heidelberg, Mannheim, Germany), Didier Meulendijks (Department of Gastrointestinal Oncology, Netherlands Cancer Institute, Amsterdam, the Netherlands), Ho-Young Yhim (Division of Hematology/Oncology, Jeonbuk National University Medical School, Jeonju, South Korea) and Franz R\"{o}del (Department of Radiotherapy and Oncology, Goethe-University, Frankfurt am Main, Germany) (providing access to illustrating dataset)

\section*{Conflict of interest}
JH and MA are supported by the Horizon 2020 Framework Programme for Research and Innovation of the European Commission, through the RISCC Network (Grant No. 847845)

\section*{Data availability statement}
The individual patient data used to illustrate our framework are not publicly available. 


\bibliographystyle{ama}
\bibliography{MyLibrary}

\clearpage

\begin{table}[t]
\caption{Sources of differences in randomized clinical trials (RCT) and observational studies with time-to-event outcomes.}
\label{tab:overview_sources}
\centering
\begin{tabular}{m{3.2cm} c|cc m{6.4cm}}
\toprule
 study s & RCT & \multicolumn{2}{c}{observational} & \\
\midrule
exposure & randomized & $e$ & $\bar{e}$ & A binary exposure is assumed in this case study.\\ 
covariate distribution   & $Z^e_s \overset{\operatorname{d}}{=} Z^{\bar{e}}_s $ & $Z^e_s$ & $Z^{\bar{e}}_s$  &  Randomization yields covariates equal in distribution within a study. \\
baseline hazard      & $\lambda_s (t)$     &   \multicolumn{2}{c}{$\lambda_s (t)$} &  May be different among studies due to variable quality of care in hospitals or unmeasured confounding.\\
treatment effect     & $\psi_s(t)$  & \multicolumn{2}{c}{$\psi_s(t)$}     &  Equal among studies (FE) or different but sampled from the same distribution (RE). \\
maximal follow-up    & $\tau_s$     & 
\multicolumn{2}{c}{$\tau_s$}        & Study specific\\
covariate distribution for censoring & $Z^c_s$ & $Z^{c,e}_s$ & $Z^{c,\bar{e}}_s$ & No unmeasured confounding for censoring is necessary to have unbiased estimators of the survival in each subgroup.\\
\bottomrule
\end{tabular}
\end{table}

\clearpage

\begin{table}[h]
\caption{Overview of events, follow-up and covariates per study and p16 exposure status. Age at time of diagnosis, tumor stage grouped into T1-2 vs T3-4, nodal stage grouped into N0 vs N1-3 and gender\cite{obermueller_prognostic_2021}. \\ \normalsize{*maximum time of observed event; Follow-up, FU; median, med; quantile q; Standard deviation, SD, tumor stage, T-stage; Nodal stage, N-stage.}} 
\scriptsize 
\centering
\begin{tabular}{llrrrrrllrrr}
\toprule
Study& p16 & n & Deaths & \multicolumn{2}{l}{FU (months)} & & Age (years) &  & T-stage & N-stage & Gender \\
 &  &  &  & med &max ($\tau$) & max* & med (q10-q90) & mean (SD) & T3/T4 & N1-3 & Female \\
\midrule
Yhim,  & 0 &  8 & 3 (38\%) & 30.0 & 63.7 & 30.8 & 59 (46-74) & 60 (12) & 3 (38\%) & 3 (38\%) & 5 (62\%) \\ 
\hspace{1em} 2011   & 1 & 39 & 9 (23\%) & 39.5 & 110.5 & 53.4 & 67 (53-76) & 66 (10) & 13 (33\%) & 15 (38\%) & 20 (51\%) \\ 
Gilbert,  & 0 & 10 & 6 (60\%) & 16.7 & 33.6 & 33.6 & 68 (54-75) & 67 (10) & 6 (60\%) & 5 (50\%) & 4 (40\%) \\ 
 \hspace{1em} 2013  & 1 & 110 & 23 (21\%) & 27.3 & 72.5 & 65.2 & 61 (46-83) & 62 (13) & 56 (51\%) & 43 (39\%) & 72 (65\%) \\
Koerber, & 0 & 15 & 6 (40\%) & 27.2 & 96.0 & 21.0 & 57 (42-80) & 58 (15) & 4 (27\%) & 5 (33\%) & 7 (47\%) \\ 
 \hspace{1em} 2014   & 1 & 75 & 17 (23\%) & 54.8 & 169.0 & 75.3 & 55 (43-82) & 58 (14) & 22 (29\%) & 17 (23\%) & 70 (93\%) \\
Mai, & 0 & 32 & 8 (25\%) & 36.5 & 205.0 & 34.0 & 60 (39-80) & 59 (14) & 7 (22\%) & 12 (38\%) & 12 (38\%) \\ 
 \hspace{1em} 2015  & 1 & 73 & 14 (19\%) & 52.0 & 201.0 & 78.0 & 58 (43-75) & 58 (12) & 20 (27\%) & 24 (33\%) & 50 (68\%) \\ 
Meulendijks, & 0 & 10 & 6 (60\%) & 25.5 & 86.0 & 34.0 & 62 (46-74) & 60 (11) & 6 (60\%) & 4 (40\%) & 2 (20\%) \\ 
 \hspace{1em} 2015   & 1 & 95 & 12 (13\%) & 35.0 & 96.0 & 45.0 & 59 (46-72) & 59 (10) & 44 (46\%) & 53 (56\%) & 54 (57\%) \\ 
Balermpas,  & 0 & 73 & 18 (25\%) & 27.2 & 325.4 & 36.3 & 62 (45-79) & 62 (12) & 26 (36\%) & 29 (40\%) & 36 (49\%) \\ 
 \hspace{1em} 2017  & 1 & 75 & 18 (24\%) & 48.8 & 186.0 & 94.1 & 58 (43-73) & 58 (12) & 17 (23\%) & 21 (28\%) & 47 (63\%) \\ 
\bottomrule
\end{tabular}
\label{tab:overview_covars}
\end{table}

\clearpage

\begin{table}[h]
\caption{Overview of different hazard ratios (HR) with their 95\% confidence interval. We show the unadjusted crude HR, the conditional HR and marginal HR where we only consider certain time periods of the standardized survival curves, applied to approach A and C. We applied artificial administrative censoring (AC) at 5 years before fitting the model. For each approach, we give the pooled HR assuming random effects (RE).}
\scriptsize
\label{tab:overview_HRs}
\begin{tabular}{@{}lllllll@{}}
\toprule
 & crude HR  & conditional HR & \multicolumn{3}{l}{marginal HR} & marginal HR\\
 &   &  & \multicolumn{3}{l}{(approach A)} & (approach C)  \\
       & & & 3 year & 5 year  & 5 year + AC & 3 year \\
       \midrule
Yhim, 2011      & 0.59 [0.16-2.17] & 0.44 [0.12-1.64]   & 0.78 [0.40-1.52] & 0.65 [0.21-1.98] & 0.67 [0.02-23]                            & 0.78 [0.49-1.25]\\
Gilbert, 2013      & 0.14 [0.05-0.37] & 0.17   [0.06-0.44] & 0.40 [0.21-0.76] & 0.23 [0.09-0.60] & 0.24 [0.00-1063]                          & 0.67 [0.52-0.87]\\
Koerber, 2014     & 0.42 [0.16-1.07] & 0.55 [0.21-1.43]   & 0.85 [0.57-1.26] & 0.75 [0.40-1.40] & 0.59 [0.01-35.75]                            & 0.84 [0.59-1.19] \\
Mai, 2015      & 0.63 [0.26-1.49] & 0.69 [0.29-1.68]   & 0.93 [0.72-1.21] & 0.90 [0.62-1.30] & 0.75 [0.00-124]                           & 0.88 [0.61-1.28]\\
Meulendijks, 2016      & 0.18 [0.07-0.48] & 0.25 [0.09-0.67]   & 0.62 [0.33-1.18] & 0.44 [0.17-1.14] & 0.41 [0.00-1677]                          & 0.68 [0.47-0.99] \\
Balermpas, 2017      & 0.79 [0.41-1.53] & 1.06 [0.55-2.05]   & 1.01 [0.79-1.29] & 1.02 [0.73-1.44] & 0.84 [0.19-3.72]                             & 1.02 [0.72-1.45] \\
\midrule
Pooled (RE) & 0.39 [0.18-0.84] & 0.47 [0.22-0.97]   & 0.82 [0.61-1.12] & 0.71 [0.42-1.20] & 0.76 [0.58-0.98]                             & 0.79 [0.66-0.94] \\ \bottomrule
\end{tabular}
\end{table}

\clearpage

\begin{table}[h]
\centering
\caption{The different approaches, with the made choices and assumptions, of standardizing survival curves, used in the framework for meta-analysis of individual patient (time-to-event) data.}
\label{tab:framework}
\begin{tabular}{@{}m{4.5cm} c CCCC l @{}} 
\toprule
\textbf{Approach}   & Population to & \multicolumn{4}{c}{Information kept} & Explainable residual variance left in\\
description & regress over & \multicolumn{4}{c}{from original study} & standardized survival curves, explained \\ 
 &  &  &  &  &  & by within or between study factors: \\
\cmidrule(lr){3-6}
 &  & $e_i$ & $Z_i$ & $\bbeta_{s(i)}$ & $\lambda_{s(i)}$ & \\
\midrule
\textbf{0} Standardizing exposure outcomes within exposure group and study population.
& \includegraphics[scale=0.6, valign=c]{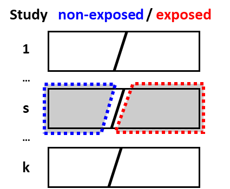} 
& \bigcheckmark & \bigcheckmark & \bigcheckmark  & \bigcheckmark 
& \splitcell{%
  \mbox{} \\[-0.2\normalbaselineskip]
 Within: measured covariate distribution \\ 
 Between: exposure effects, measured \\
covariate distribution and center specific \\
baseline effects.
  \\[-0.2\normalbaselineskip] \mbox{}%
} \\ 

\textbf{A} Balancing study covariate distribution between exposure groups within the study.
& \includegraphics[scale=0.6, valign=c]{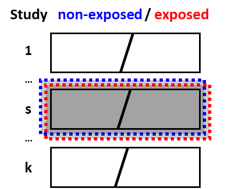} 
&  & \bigcheckmark & & \bigcheckmark 
& \splitcell{%
Within: / \\ 
Between: exposure effects, measured \\
covariate distribution and center specific \\
baseline effects.
} \\ 

\textbf{B} Case-mix adjusted standardized survival curves, a direct standardization using full observed and measured covariate distribution across all studies and exposure groups.
& \includegraphics[scale=0.6, valign=c]{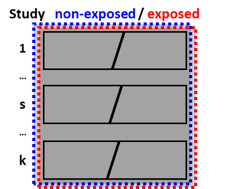} 
&  & \bigcheckmark & & 
& \splitcell{%
 Within: / \\ 
 Between: exposure effects and \\
 center specific baseline effects.
} \\ 

\textbf{C} Case-mix adjusted standardized survival curves, Standardizing exposure outcomes within the full observed patient distribution and using the study s specific exposure effect but keeping the observed individual baseline hazard.
& \includegraphics[scale=0.6,valign=c]{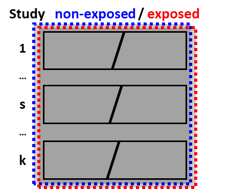} 
&  & \bigcheckmark & \bigcheckmark & \bigcheckmark  
& \splitcell{%
 Within: / \\ 
 Between: exposure effects
} \\ 

\textbf{D} Standardizing over a full observed and measured exposure specific covariate distribution (here illustrated with the exposed population) across all studies
& \includegraphics[width=2.3cm, valign=c]{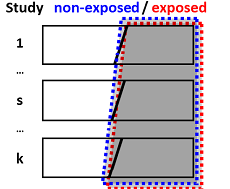} &  & \bigcheckmark &  &   
& \splitcell{%
 Within: / \\
 Between: exposure effects and \\
 center specific baseline effects.
} \\ 
\bottomrule
\end{tabular}
\end{table}

\clearpage

\counterwithin{figure}{section}
\counterwithin{table}{section}
\renewcommand\thefigure{\thesection\arabic{figure}}
\renewcommand\thetable{\thesection\arabic{table}}

\appendix

\addcontentsline{toc}{section}{Supporting information}
\section*{Supporting information}
\section{Results of different approaches for the different estimands, using different time periods and potential artificial administrative censoring}

\begin{figure}[ht]
\centering
\includegraphics[width=0.99\textwidth]{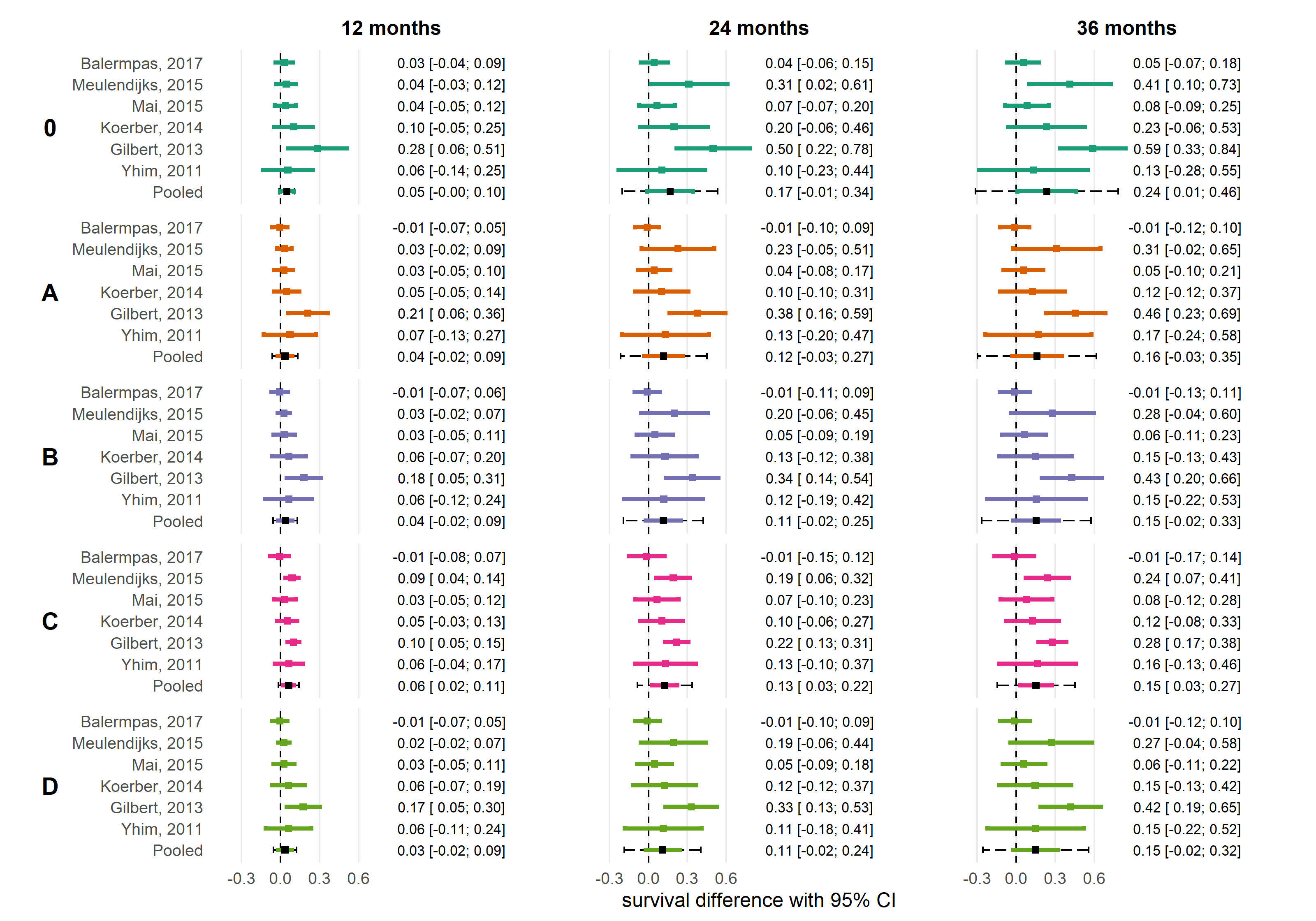}
\caption{Forest plots of survival difference, reflecting the contrast in survival between p16+ and p16- anal cancer, at 12, 24 and 36 months using standardized survival curves with different approaches, using a flexible proportional hazards model. Pooled for each approach and time assuming random effects, the prediction interval is shown with the black dotted lines. The standard error for each survival difference is obtained using the jackknife method.}
\label{fig:forest_survdif_uncensored}
\end{figure}

\begin{figure}[ht]
\centering
\includegraphics[width=0.99\textwidth]{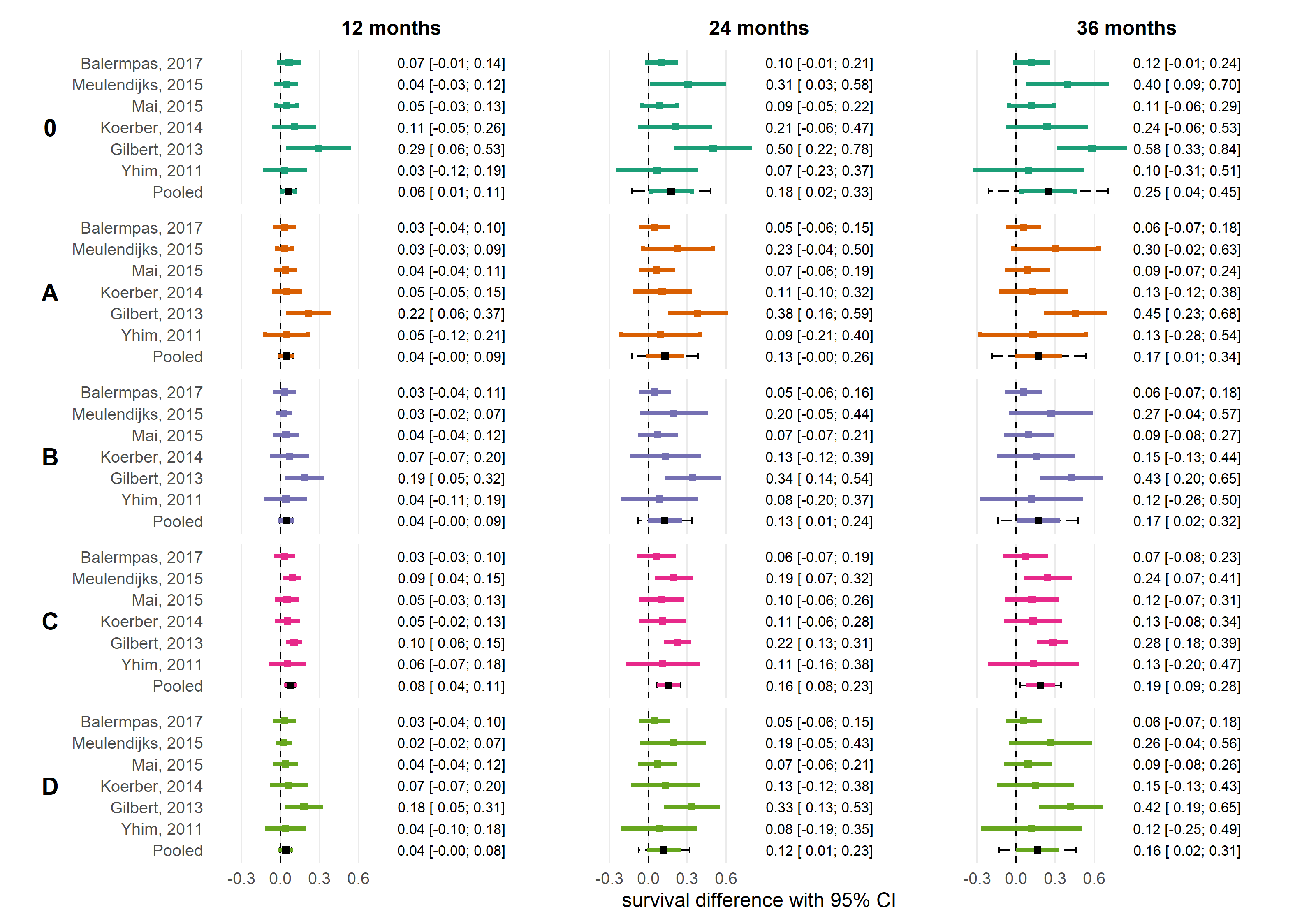}
\caption{Forest plots of survival difference, reflecting the contrast in survival between p16+ and p16- anal cancer, at 12, 24 and 36 months using standardized survival curves with different approaches, using a flexible proportional hazards model but in which we censored the patients artificially at 60 months before fitting the model. Pooled for each approach and time assuming random effects, the prediction interval shown with the black dotted lines. The standard error for each survival difference is obtained using the jackknife method.}
\label{fig:forest_survdif_censored}
\end{figure}

\begin{figure}[ht]
\centering
\includegraphics[width=0.99\textwidth]{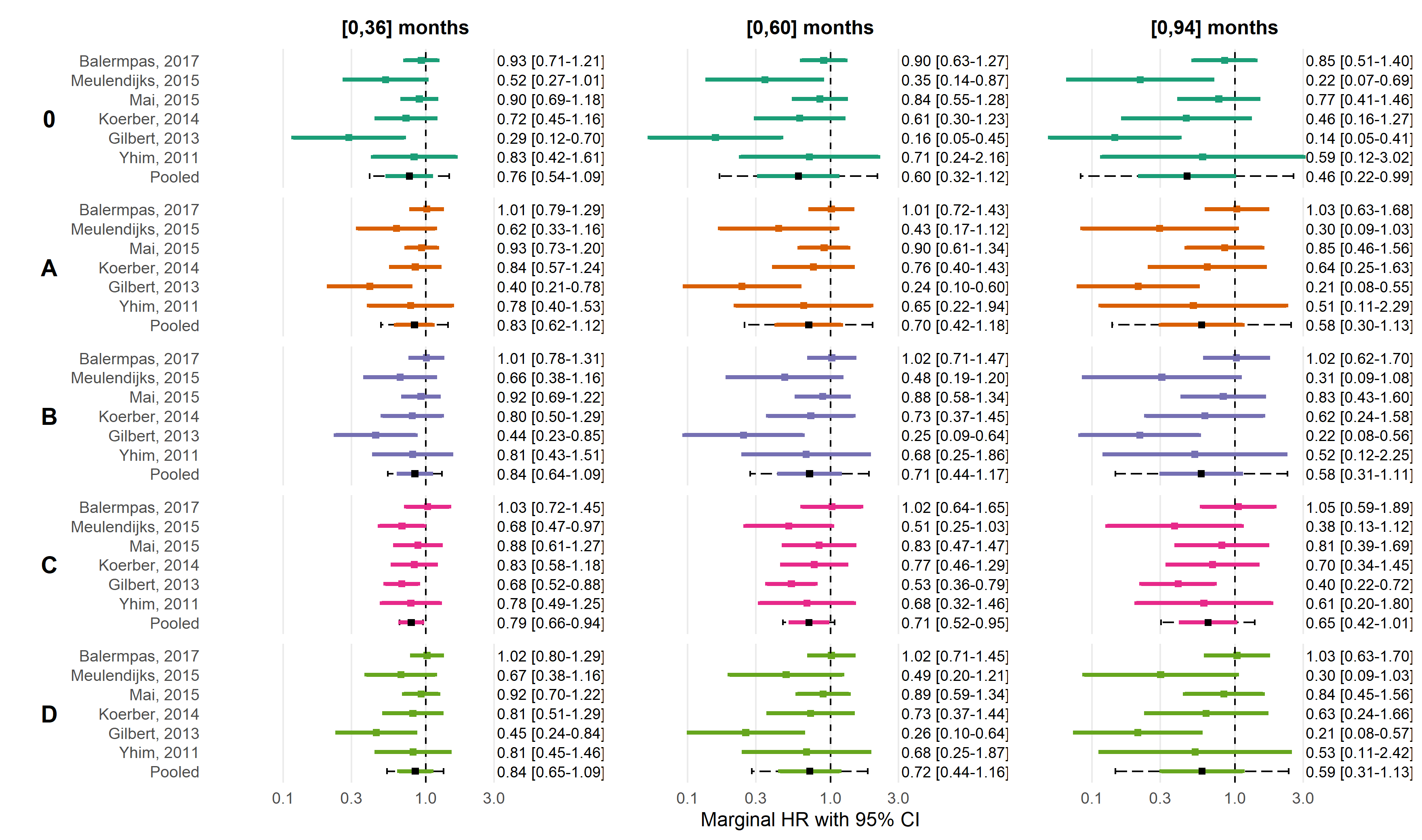}
\caption{Forest plots with random effects meta-analysis of marginal hazard ratios (HR), reflecting the contrast in survival between p16+ and p16- anal cancer, after different standardization approaches and for different time periods considered. All studies had a follow-up (FU) of 36 months, but not all studies had a FU of 60 months and the standardized curves are thus an extrapolation. The last observed event occurred at 94 months. Pooled for each approach and time assuming random effects, the prediction interval is shown with the black dotted lines. Standard errors were obtained using jackknife resampling.}
\label{fig:forest_margHR_uncensored}
\end{figure}

\begin{figure}[ht]
\centering
\includegraphics[width=0.88\textwidth]{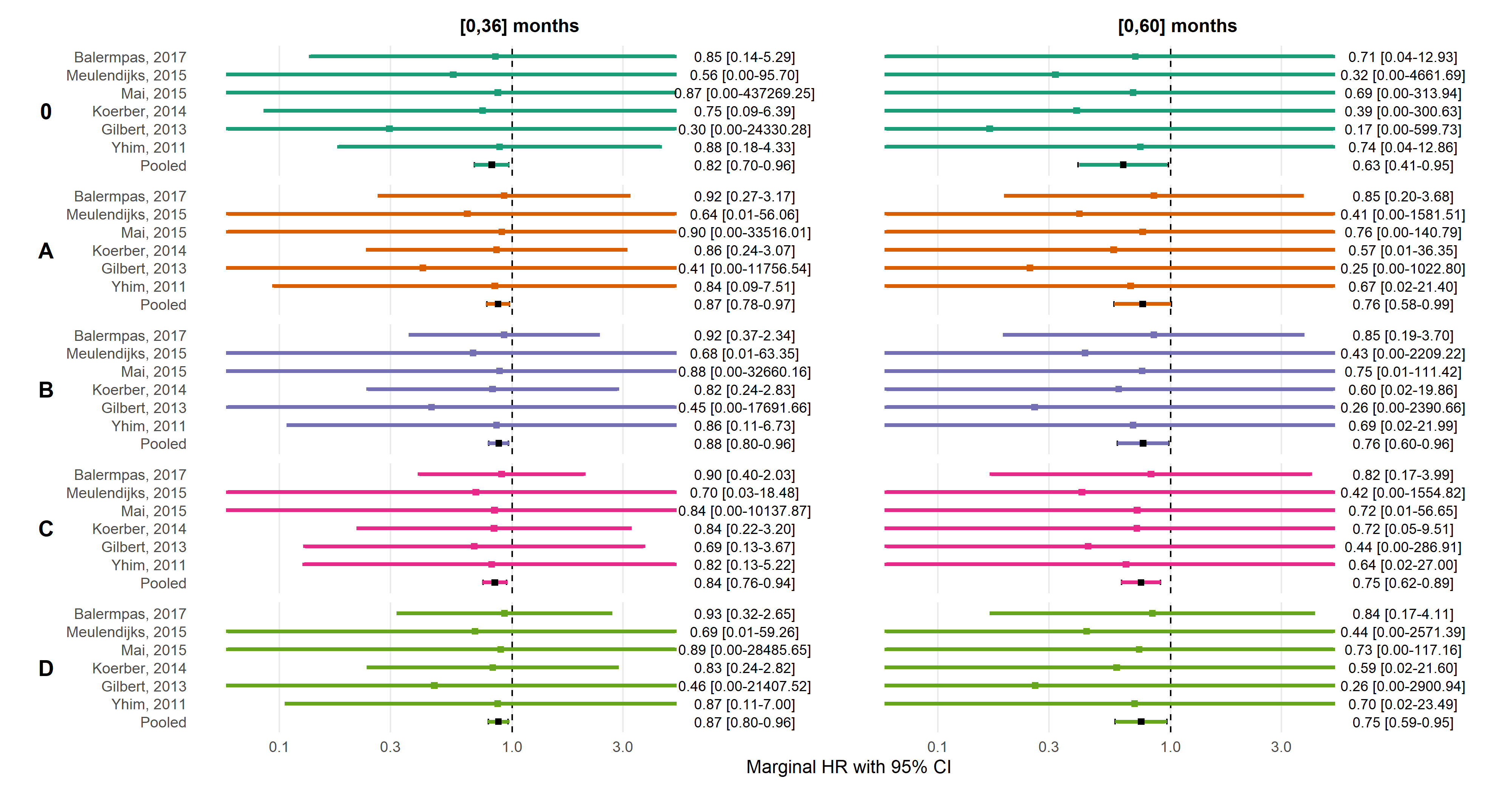}
\caption{Forest plots with random effects meta-analysis of marginal hazard ratios (HR), reflecting the contrast in survival between p16+ and p16- anal cancer, after different standardization approaches and for different time periods similar as figure \ref{fig:forest_margHR_uncensored} but with artificial administrative censoring at 60 months. So all observations were censored at 60 months before fitting the model. Pooled for each approach and time assuming random effects, the prediction interval is shown with the black dotted lines. Standard errors were obtained using jackknife resampling.}
\label{fig:forest_margHR_censored}
\end{figure}

\clearpage

\section{Difference between manual approach and rstpm2 package}

\begin{table}[ht]
\centering
\caption{Comparison between jackknife standard error (se) and the se of the rstpm package. The rstpm package does not automatically respect the hazard constraint, we manually replace any negative point estimates with the lowest observed (non zero) hazard. For most studies and approaches the difference between the two methods is minimal, the Yhim study and Meulendijks study show the largest differences between the se of both methods with jackknife having 33\% larger se than the se of the rstpm2 package. With approach C,  the comparison is more difficult as rstpm2 had negative hazards which infects now all studies as the baseline hazard for the Meulendijks study is used in all study effect differences}
\begin{tabular}{rllll}
  \toprule
 & \multicolumn{4}{l}{risk difference at 36 months [95\% CI] standard error} \\
  \midrule
  rstpm2 & A & B & C & D \\ 
  \midrule
Yhim, 2011 &  0.17 [-0.14; 0.48] 0.16 &  0.16 [-0.13; 0.44] 0.14 &  0.16 [-0.06; 0.38] 0.11 &  0.17 [-0.14; 0.48] 0.16 \\ 
Gilbert, 2013 &  0.46 [ 0.20; 0.71] 0.13 &  0.43 [ 0.17; 0.69] 0.13 &  0.28 [ 0.17; 0.39] 0.06 &  0.45 [ 0.19; 0.71] 0.13 \\ 
Koerber, 2014 &  0.12 [-0.09; 0.34] 0.11 &  0.15 [-0.10; 0.40] 0.13 &  0.12 [-0.06; 0.31] 0.09 &  0.16 [-0.10; 0.42] 0.13 \\ 
Mai, 2015 &  0.06 [-0.08; 0.19] 0.07 &  0.06 [-0.09; 0.22] 0.08 &  0.08 [-0.10; 0.26] 0.09 &  0.07 [-0.10; 0.24] 0.09 \\ 
Meulendijks, 2015 &  0.31 [ 0.06; 0.57] 0.13 &  0.28 [ 0.04; 0.51] 0.12 &  0.24 [ 0.10; 0.38] 0.07 &  0.30 [ 0.05; 0.55] 0.13 \\ 
Balermpas, 2017 & -0.01 [-0.13; 0.10] 0.06 & -0.01 [-0.13; 0.11] 0.06 & -0.01 [-0.17; 0.14] 0.08 & -0.01 [-0.14; 0.12] 0.07 \\ 
  \midrule
  jackknife & A & B & C & D \\ 
  \midrule
Yhim, 2011 &  0.17 [-0.24; 0.58] 0.21 &  0.15 [-0.22; 0.53] 0.19 &  0.16 [-0.13; 0.46] 0.15 &  0.15 [-0.22; 0.52] 0.19 \\ 
Gilbert, 2013 &  0.46 [ 0.23; 0.69] 0.12 &  0.43 [ 0.20; 0.66] 0.12 &  0.28 [ 0.17; 0.38] 0.05 &  0.42 [ 0.19; 0.65] 0.12 \\ 
Koerber, 2014 &  0.12 [-0.12; 0.37] 0.13 &  0.15 [-0.13; 0.43] 0.14 &  0.12 [-0.08; 0.33] 0.10 &  0.15 [-0.13; 0.42] 0.14 \\ 
Mai, 2015 &  0.05 [-0.10; 0.21] 0.08 &  0.06 [-0.11; 0.23] 0.09 &  0.08 [-0.12; 0.28] 0.10 &  0.06 [-0.11; 0.22] 0.08 \\ 
Meulendijks, 2015 &  0.31 [-0.02; 0.65] 0.17 &  0.28 [-0.04; 0.60] 0.16 &  0.24 [ 0.07; 0.41] 0.08 &  0.27 [-0.04; 0.58] 0.16 \\ 
Balermpas, 2017 & -0.01 [-0.12; 0.10] 0.06 & -0.01 [-0.13; 0.11] 0.06 & -0.01 [-0.17; 0.14] 0.08 & -0.01 [-0.12; 0.10] 0.06 \\ 
\bottomrule
\end{tabular}
\end{table}

\clearpage

\clearpage

\section{Transportability checks}

\begin{figure}[h]
\centering
\includegraphics[width=0.59\textwidth]{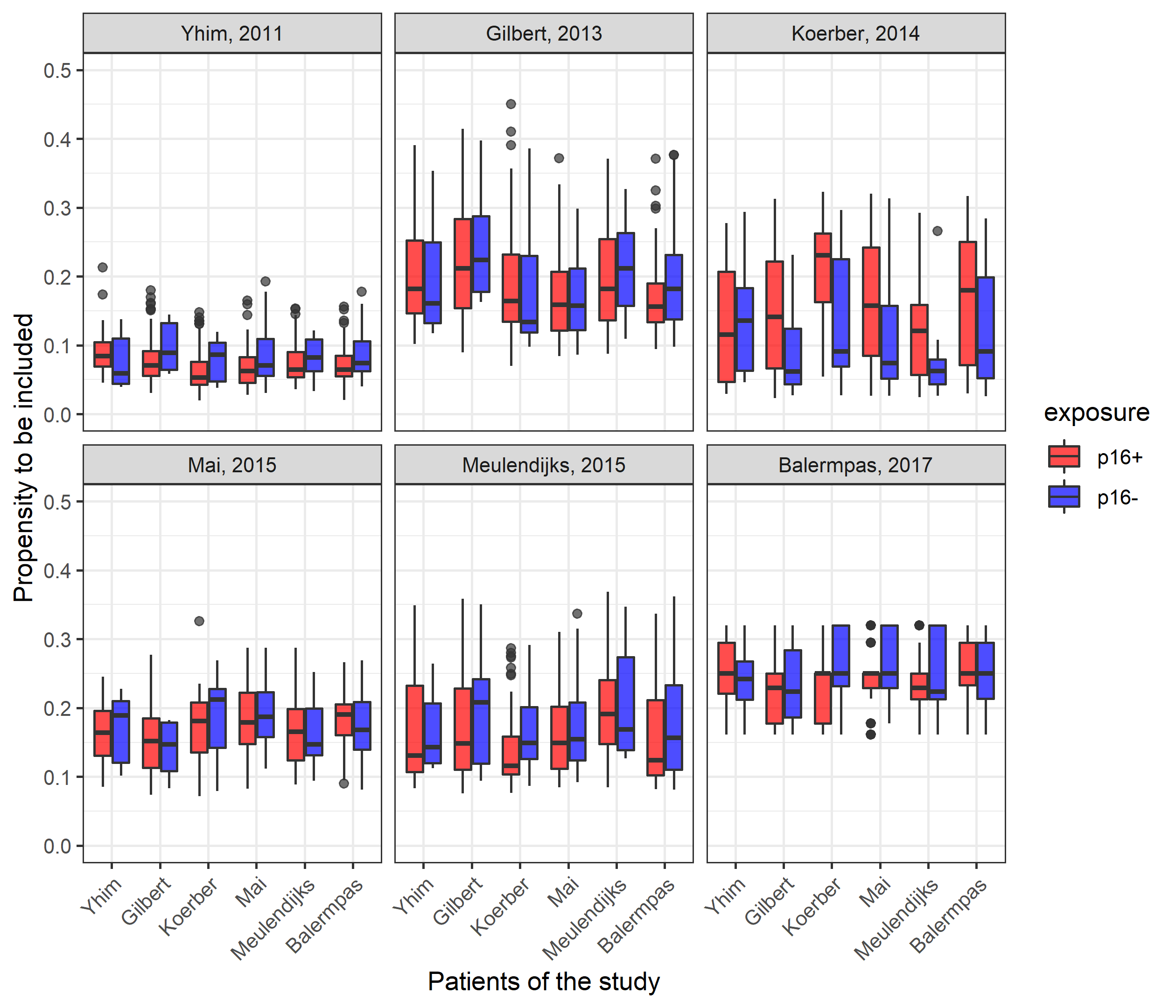}
\caption{Propensity to be included in the different studies per exposure and study subgroup.}
\label{fig:prop_subgroup}
\end{figure}

\end{document}